\documentclass[prl,twocolumn,showpacs]{revtex4}
\usepackage{amsmath}
\usepackage{dcolumn}
\usepackage{graphicx}
\usepackage{bm}

\begin{document}

\title{Maximized string order parameters in the valence bond solid states of
quantum integer spin chains}
\author{Hong-Hao Tu$^{1}$}
\author{Guang-Ming Zhang$^{1}$}
\email{gmzhang@tsinghua.edu.cn}
\author{Tao Xiang$^{2,3}$}
\affiliation{$^{1}$Department of Physics, Tsinghua University, Beijing 100084, China}
\affiliation{$^{2}$Institute of Physics, Chinese Academy of Sciences, Beijing 100190,
China}
\affiliation{$^{3}$Institute of Theoretical Physics, Chinese Academy of Sciences, Beijing
100190, China }
\date{\today }

\begin{abstract}
We propose a set of maximized string order parameters to describe the hidden
topological order in the valence bond solid states of quantum integer spin-$%
S $ chains. These optimized string order parameters involve spin-twist
angles corresponding to $Z_{S+1}$ rotations around $z$ or $x$ axes,
suggesting a hidden $Z_{S+1}\times Z_{S+1}$ symmetry. Our results also
suggest that a local triplet excitation in the valence bond solid states
carries a $Z_{S+1}$ topological charge measured by these maximized string
order parameters.
\end{abstract}

\pacs{03.67.-a, 75.10.Pq, 64.70.Tg, 75.10.Jm}
\maketitle

Determining order parameters is one of the most important issues in the
study of strongly correlated systems. It is the basis of Landau theory in
describing continuous phase transitions between different phases with
spontaneous symmetry breaking. However, some novel phenomena or phases, for
example, the fractional quantum Hall states and the Haldane gap phenomena in
quantum integer-spin chains, are not amenable to such a\ description. In
particular, to find an order parameter in a system with strong frustrations
or quantum fluctuations is challenging and highly nontrivial \cite%
{Oshikawa-2006,Cirac-2008}.

For quantum integer-spin antiferromagnetic Heisenberg chains, Haldane
predicted that there is a finite excitation gap above the ground state \cite%
{Haldane-1983}. To understand this intriguing conjecture, Affleck, Kennedy,
Lieb, and Tasaki (AKLT) proposed a set of exactly solvable quantum spin
models whose ground states are the valence bond solid (VBS) states \cite%
{AKLT}. In the $S=1$ VBS state, there is a hidden antiferromagnetic order
which can be characterized by string order parameters (SOPs) \cite{den
Nijs-1989}. Corresponding to this nonlocal string order, Kennedy and Tasaki
showed that there is a hidden $Z_{2}\times Z_{2}$ symmetry \cite%
{Kennedy-1991}. To reveal this symmetry clearly, they introduced a nonlocal
unitary transformation to convert the VBS state to a \textquotedblleft
diluted\textquotedblright\ ferromagnetic state and showed that the ground
state is fourfold degenerate if this symmetry is broken in an open chain.

For quantum integer-spin chains with $S>1$, it is extremely difficult to
find the corresponding nonlocal unitary transformation to convert the VBS
state into a long-range ordered state. So far no one has succeeded along
this direction. However, this is not the only approach for revealing a
hidden topological symmetry. This hidden symmetry can be also identified
from the non-local string order parameters of the ground state. So the
extended SOPs with a twist angle $\theta =\pi /S$ were proposed to account
for the hidden order in the VBS states \cite{Oshikawa-1992,Suzuki-1995}.
Actually, this kind of extended SOPs does not give a comprehensive
description to the Haldane phase in these systems \cite{Schollwock-1996}.
Recently, we have extended the $S=1$ VBS (which is $SO(3)$ invariant) to the
$SO(2n+1)$ symmetric system \cite{HHTu-2008}. We find that these $SO(2n+1)$
symmetric matrix product states are the exact ground states of Hamiltonians
of spin $S=n$ with either $SO(2n+1)$ or $SU(2)$ symmetry. Furthermore, we
show that this system possesses a hidden topological $(Z_{2}\times
Z_{2})^{n} $ symmetry and the ground state can be characterized by $n$-set
of hidden antiferromagnetic orders. This study suggests that the topological
order is generally more complicated than the $\theta $-twisted SOP as
previously studied in the VBS states. It motivates us to consider how to
find proper SOPs to characterize the hidden topological order in the higher
integer-spin VBS states \cite{Nachtergaele-Hadley}.

In this paper, we will propose a systematical method to find the optimized
SOPs in the VBS states. By generalizing the $S=1$ den Nijs-Rommelse SOP to
arbitrary integer spin-$S$ VBS states, the corresponding hidden long-range
order can be most comprehensively manifested by the maximal string
correlation functions between two polynomial spin operators%
\begin{equation}
A_{j}^{\alpha }=\sum_{n=0}^{S}a_{n}\left( S_{j}^{\alpha }\right) ^{n},\qquad
(\alpha =x,y,z),  \label{eq:A}
\end{equation}%
where $a_{n}$ are the coefficients which maximize the string correlation
function. We find that these maximized SOPs suggest a hidden $Z_{S+1}\times
Z_{S+1}$ symmetry, in consistent with the degeneracy of the ground states in
an open chain case. Moreover, our optimized SOPs measure a $Z_{S+1}$
topological charge carried by a localized triplet excitation in the VBS
state.

The AKLT Hamiltonian is defined by \cite{AKLT,Arovas-1988}
\begin{equation}
H_{\mathrm{AKLT}}=\sum_{i}\sum_{J=S+1}^{2S}K_{J}\mathcal{P}_{J}(i,i+1),
\label{VBS}
\end{equation}%
where $K_{J}>0$ and $\mathcal{P}_{J}(i,i+1)$ is to project two spins at $i$
and $i+1$ onto the subspace of the total spin $J$. This Hamiltonian can be
written as a polynomial of $SU(2)$ invariant nearest-neighbor spin exchange
interactions. The VBS state is the ground state of this Hamiltonian.

In the Schwinger boson representation, the spin operators are expressed by $%
S_{i}^{+}=a_{i}^{\dagger }b_{i}$, $S_{i}^{-}=b_{i}^{\dagger }a_{i}$, $%
S_{i}^{z}=(a_{i}^{\dagger }a_{i}-b_{i}^{\dagger }b_{i})/2$ with a local
constraint $a_{i}^{\dagger }a_{i}+b_{i}^{\dagger }b_{i}=2S$. The VBS state
in a length-$L$\ periodic chain is then given by
\begin{equation}
\left| \mathrm{VBS}\right\rangle =\prod_{i=1}^{L}(a_{i}^{\dagger
}b_{i+1}^{\dagger }-b_{i}^{\dagger }a_{i+1}^{\dagger })^{S}\left| \mathrm{vac%
}\right\rangle .
\end{equation}%
In an open chain, the VBS states have effectively two spin-$S/2$ edge states
at the two ends of the chain. In the thermodynamic limit, $L\rightarrow
\infty $, the VBS states with different edge states are asymptotically
orthogonal to each other, giving rise to $(S+1)^{2}$-fold degenerate ground
states.

The VBS state can be also represented in a matrix product form \cite%
{Fannes-1989,Klumper-1991,Suzuki-1995}. By using the binomial theorem, it
can be shown that
\begin{equation}
\left| \mathrm{VBS}\right\rangle =\sum_{\{m_{i}\}=-S}^{S}\mathrm{Tr}%
(B^{[m_{1}]}\cdots B^{[m_{L}]})\left| m_{1}\cdots m_{L}\right\rangle ,
\end{equation}%
where $B^{[m]}$ is a $(S+1)\times (S+1)$ matrix defined by its matrix
elements as
\begin{eqnarray}
B^{[m]}(p,q) &=&(-1)^{S-p+1}\sqrt{(S+m)!(S-m)!}  \notag \\
&&\times \sqrt{\binom{S}{p-1}\binom{S}{q-1}}\delta _{m,q-p},
\end{eqnarray}%
where $1\leq p$, $q\leq S+1.$

The correlation function of the VBS state in the matrix product form can be
evaluated using the transfer matrix method \cite{Klumper-1991,Suzuki-1995}.
To do this, let us first introduce the following $(S+1)^{2}\times (S+1)^{2}$
transfer matrix:
\begin{equation}
G_{P}=\sum_{m,m^{\prime }}\left\langle m^{\prime }\right| \hat{P}\left|
m\right\rangle \left( \bar{B}^{[m^{\prime }]}\otimes B^{[m]}\right) ,
\label{eq:TranMat}
\end{equation}%
where $\hat{P}$ is a local operator acting on a single site, and $\bar{B}$
denotes the complex conjugate of $B$. For the identity operator $\hat{P}=I$,
$G_{P}$ is a Hermitian transfer matrix $G=\sum_{m}\left( \bar{B}%
^{[m]}\otimes B^{[m]}\right) $. With these definitions, it is
straightforward to show that the two-point spin correlation function can be
expressed as
\begin{equation}
\left\langle S_{i}^{z}S_{j}^{z}\right\rangle =\lim_{L\rightarrow \infty }%
\frac{\mathrm{Tr}[(G)^{L-j+i-1}G_{S}(G)^{j-i-1}G_{S}]}{\mathrm{Tr}(G)^{L}},
\end{equation}%
where
\begin{equation*}
G_{S}=\sum_{m}m\left( \bar{B}^{[m]}\otimes B^{[m]}\right) .
\end{equation*}%
In the long distance limit, the two-point spin correlation function always
decays exponentially with the distance between the sites $i$ and $j$,
\begin{equation}
\left\langle S_{i}^{z}S_{j}^{z}\right\rangle \sim \exp \left( -\frac{|j-i|}{%
\xi }\right) ,
\end{equation}%
where $\xi =1/\ln (1+2/S)$ is the correlation length \cite{Arovas-1988}.

In order to describe the hidden topological order in the VBS states, let us
introduce the following generalized string correlation function \cite%
{Suzuki-1995}:
\begin{equation}
\mathcal{O}_{A}^{\alpha }(\theta )=\lim_{\left\vert j-i\right\vert
\rightarrow \infty }\langle \left( A_{i}^{\alpha }\right) ^{\dagger
}\prod_{k=i}^{j-1}e^{i\theta S_{k}^{\alpha }}A_{j}^{\alpha }\rangle ,
\label{G-String}
\end{equation}%
where $A_{j}^{\alpha }$ is defined by Eq.~(\ref{eq:A}). The value of $a_{0}$
in $A_{j}^{\alpha }$ can be fixed by demanding the expectation value of $%
A_{j}^{\alpha }$ to be zero, i.e., $a_{0}=-\sum_{n=1}^{S}a_{n}\left\langle
(S_{j}^{\alpha })^{n}\right\rangle $.

Since the VBS state is spin SU(2) rotational invariant, we need only to
evaluate the $z$-component of the string correlation function $\mathcal{O}%
_{A}^{z}(\theta )$. Based on the transfer matrix technique, it can be shown
that
\begin{equation}
\mathcal{O}_{A}^{z}(\theta )=\lim_{\left| j-i\right| \rightarrow \infty
}\lim_{L\rightarrow \infty }\frac{\text{Tr}[(G)^{L-j+i-1}G_{A^{\dagger }}(G_{%
\mathcal{O}})^{j-i-1}G_{A}]}{\text{Tr}(G)^{L}},  \label{eq:CompSOP}
\end{equation}%
where $G_{\mathcal{O}}$, $G_{A^{\dagger }}$, and $G_{A}$ are obtained by
replacing operator $\hat{P}$ in Eq. (\ref{eq:TranMat}) by $\exp (i\theta
S^{z})$, $\left( A^{z}\right) ^{\dagger }\exp (i\theta S^{z})$, and $A^{z}$,
respectively. We emphasize that only $G$ and $G_{\mathcal{O}}$ are
Hermitian, while $G_{A^{\dagger }}$ and $G_{A}$ are not.

In the limit of $|j-i|\rightarrow \infty $ and $L\rightarrow \infty $, $%
\mathcal{O}_{A}^{z}(\theta )$ is determined purely by the largest
eigenvalues and eigenvectors of $G$ and $G_{\mathcal{O}}$. We find that Eq.~(%
\ref{eq:CompSOP}) can be simplified as%
\begin{equation}
\mathcal{O}_{A}^{z}(\theta )=\frac{1}{\lambda _{\max }^{2}}\left|
\left\langle \lambda _{\max }^{\mathcal{O}}\right| G_{A}\left| \lambda
_{\max }\right\rangle \right| ^{2}.  \label{StringOrder}
\end{equation}%
where $|\lambda _{\mathrm{max}}\rangle $ and $|\lambda _{\mathrm{max}}^{%
\mathcal{O}}\rangle $ are the eigenvectors corresponding to the same largest
eigenvalue of $G$ and $G_{\mathcal{O}}$, respectively.

When $\theta =0$, $\mathcal{O}_{A}^{z}(\theta )$ becomes the ordinary
two-point correlation function. It vanishes because $|\lambda _{\max }^{%
\mathcal{O}}\rangle =|\lambda _{\max }\rangle $ for $\theta =0$ and the
vector $G_{A}\left| \lambda _{\max }\right\rangle $ is orthogonal to $\left|
\lambda _{\max }\right\rangle $. If we fix the form of the operator $A^{z}$
and tune the spin-twist angle $\theta $, $\mathcal{O}_{A}^{z}(\theta )$ will
reach its maximum only when the vector $G_{A}\left| \lambda _{\max
}\right\rangle $ is parallel to the vector $|\lambda _{\max }^{\mathcal{O}%
}\rangle $. This means that in order to maximize the SOP, $|\lambda _{\max
}^{\mathcal{O}}\rangle $ is orthogonal to $\left| \lambda _{\max
}\right\rangle $, i.e., $\left\langle \lambda _{\max }^{\mathcal{O}}|\lambda
_{\max }\right\rangle =0$. For the VBS state, the eigenvectors $|\lambda
_{\max }\rangle $ and $|\lambda _{\max }^{\mathcal{O}}\rangle $ can be
readily calculated, we find that the orthogonality condition can be
satisfied \textit{if and only if} the following equation is satisfied
\begin{equation}
1+e^{i\theta }+e^{i2\theta }+...+e^{iS\theta }=0.
\end{equation}%
Thus, the spin-twist angle corresponding to the maximal SOP is determined by
\begin{equation}
\theta =\frac{2n\pi }{S+1},\qquad (n=1,\cdots ,S).  \label{rotation}
\end{equation}%
Importantly, these spin-twist angles correspond to a discrete symmetry group
$Z_{S+1}$. Because of the spin SU(2) rotational symmetry, a similar maximal
SOP can be found in terms of the $x$-component operators $\mathcal{O}%
_{A}^{x}(\theta )$. The spin-twist angles corresponding to the maximal $%
\mathcal{O}_{A}^{x}(\theta )$ are also given by Eq.(\ref{rotation}).

In the conventional Landau theory, it is well-known that the maximized order
parameters fully characterize the symmetry of the low-temperature ordered
phases, and these order parameters can be also used to describe the possible
phase transitions as decreasing the temperature from the high temperature
disordered phases. When we generalize the similar rule to the present case,
the maximal SOPs fully describe the hidden long-range order of the spin-$S$
VBS states, suggesting that there exists a hidden $Z_{S+1}\times Z_{S+1}$
symmetry. In a finite open chain, this hidden $Z_{S+1}\times Z_{S+1}$
symmetry is broken in the ground states, leading to the spin-$\frac{S}{2}$
edge states with $(S+1)^{2}$-fold degeneracy. We would like to emphasize
that these spin-$\frac{S}{2}$ edge states are dictated by the underlying
low-energy effective field theory, i.e., $O(3)$ nonlinear sigma model plus a
topological term \cite{TKNg-1994}. Moreover, the hidden $Z_{S+1}\times
Z_{S+1}$ symmetry also incorporates the previous results given by Oshikawa %
\cite{Oshikawa-1992}. For an odd integer $S$ chain, the $Z_{S+1}$ group has
always a $Z_{2}$ subgroup with $\theta =\pi $. However, for an even integer $%
S$ chain, a twist angle $\theta =\pi $ is always absent. This result
explains why the hidden $Z_{2}\times Z_{2}$ symmetry described by the den
Nijs-Rommelse SOP is broken in the odd-$S$ VBS states but not in the even-$S$
ones \cite{Oshikawa-1992}.

Let us consider the hidden topological symmetries in the first few cases of
the spin-$S$ VBS states. For the $S=1$ VBS state, we have $%
A_{j}^{z}=S_{j}^{z}$. The maximal eigenvalues of the transfer matrices $G$
and $G_{\mathcal{O}}$ are equal to $\lambda _{\max }=3$, and their
eigenvectors are given by
\begin{eqnarray*}
\left| \lambda _{\max }\right\rangle  &=&\frac{1}{\sqrt{2}}\left(
\begin{array}{cccc}
1 & 0 & 0 & 1%
\end{array}%
\right) ^{T}, \\
\left| \lambda _{\max }^{\mathcal{O}}\right\rangle  &=&\frac{1}{\sqrt{2}}%
\left(
\begin{array}{cccc}
e^{i\theta } & 0 & 0 & 1%
\end{array}%
\right) ^{T},
\end{eqnarray*}%
respectively. With a simple calculation, it can be shown that $\mathcal{O}%
_{A}^{z}(\theta )=(4/9)\sin ^{2}(\theta /2)$, from which its maximal value
corresponds to $\theta =\pi $. This result is fully consistent with the
orthogonality condition. According to the nonlocal unitary transformation,
the corresponding VBS state does exhibits a $Z_{2}\times Z_{2}$ topological
symmetry. Therefore, it is a reliable method of using the maximal SOPs to
reveal the hidden symmetry.

Next let us consider the $S=2$ VBS state, the maximal eigenvalues of the
transfer matrices $G$ and $G_{\mathcal{O}}$ is $\lambda _{\max }=40$ and the
corresponding eigenvectors are given by
\begin{eqnarray*}
\left| \lambda _{\max }\right\rangle &=&\frac{1}{\sqrt{3}}%
\begin{pmatrix}
1 & 0 & 0 & 0 & 1 & 0 & 0 & 0 & 1%
\end{pmatrix}%
^{T}, \\
\left| \lambda _{\max }^{\mathcal{O}}\right\rangle &=&\frac{1}{\sqrt{3}}%
\begin{pmatrix}
e^{2i\theta } & 0 & 0 & 0 & e^{i\theta } & 0 & 0 & 0 & 1%
\end{pmatrix}%
^{T}.
\end{eqnarray*}%
When $\theta =2\pi /3$ or $\theta =4\pi /3$, the SOP (\ref{G-String}) is
maximal. From the condition that $G_{A}\left| \lambda _{\max }\right\rangle $
is parallel to $\left| \lambda _{\max }^{\mathcal{O}}\right\rangle $, we can
find the corresponding optimal operators displaying the maximal string
correlation function as the form
\begin{equation}
A_{j}^{z}=S_{j}^{z}\pm \frac{5\sqrt{3}i}{7}\left[ (S_{j}^{z})^{2}-2\right] ,
\end{equation}%
for $\theta =2\pi /3$ and $\theta =4\pi /3$, respectively. The maximal value
of the SOPs are $\mathcal{O}_{A}^{z}\left( \theta \right) =3$. In this
nontrivial case, it is the linear combination of spin operator $S_{j}^{z}$
and its spin quadrupole operator $(S_{j}^{z})^{2}-2$ that exhibits the
maximal hidden string order instead of the spin operator itself. The hidden
topological symmetry in the $S=2$ VBS state corresponds to the $Z_{3}\times
Z_{3}$ discrete symmetry.

For the $S=3$ case, the spin-twist angles for the maximal SOPs are
determined by $\theta =\pi /2$, $\pi $, $3\pi /2$ from the orthogonality
condition. Similar to the previous procedure, it is straightforward to show
that the corresponding combinations of the spin operators are given by
\begin{equation}
A_{j}^{z}=S_{j}^{z}\pm \frac{15i}{67}\left[ (S_{j}^{z})^{2}-4\right] -\frac{7%
}{67}(S_{j}^{z})^{3}
\end{equation}%
for $\theta =\pi /2$ or $3\pi /2$, and
\begin{equation}
A_{j}^{z}=\frac{89\sqrt{2}}{67}\left[ S_{j}^{z}-\frac{14}{89}(S_{j}^{z})^{3}%
\right]
\end{equation}%
for $\theta =\pi $. The maximal value of the SOPs is $\mathcal{O}%
_{A}^{z}\left( \theta \right) =2592/4489$, and the corresponding hidden
symmetry is $Z_{4}\times Z_{4}$.

Accordingly, the above discussions can be readily extended to any higher
integer spin-$S$ VBS states. In addition to the VBS states, it can be also
generalized to reveal the possible hidden topological order of arbitrary
matrix product states. Thus, this is a systematic approach to analyze the
hidden topological symmetry of matrix product states.

\begin{figure}[tbp]
\includegraphics[width=\hsize]{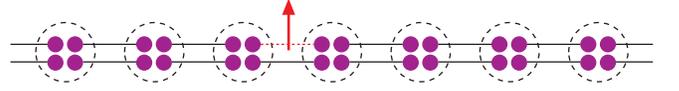}
\caption{(Color online) The schematic of a crackion excitation in a spin-$2$
VBS state. The solid lines represent the valence bond singlets. A dot
represents a spin-1/2 Schwinger boson. The four bosons enclosed by each
dashed circle form a spin-$2$ state. The arrow represents a crackion of
local triplet defect.}
\label{fg:Crackion}
\end{figure}

In fact, the topological property of the ground state can be also understood
from its elementary excitations. For the AKLT model, to create an elementary
excitation is to insert a triplet defect in the VBS ground state. This kind
of excitation is called crackion \cite{Fath-1993}. Under the Kennedy-Tasaki
unitary transformation for $S=1$, a crackion is a kink linking two
ferromagnetic ordered ground states from left to right. In higher-$S$
systems, although the Kennedy-Tasaki unitary transformation does not
transfer the VBS state to a ferromagnetic one, the kink or soliton-like
feature of crackion can still be revealed by studying the string correlation
function in the presence of a triplet defect \cite{Suzuki-1995}. An example
of a crackion in the $S=2$ VBS state is depicted in Fig.~\ref{fg:Crackion}.

With the Schwinger boson representation, the wave function of a localized
crackion between $k$ and $k+1$ sites can be constructed by replacing a
singlet $(a_{k}^{\dagger }b_{k+1}^{\dagger }-b_{k}^{\dagger
}a_{k+1}^{\dagger })$ with one of the triplet operators $T^{a}$
\begin{equation*}
T^{1}=a_{k}^{\dagger }a_{k+1}^{\dagger },\text{ }T^{0}=a_{k}^{\dagger
}b_{k+1}^{\dagger }+b_{k}^{\dagger }a_{k+1}^{\dagger },\text{ }%
T^{-1}=b_{k}^{\dagger }b_{k+1}^{\dagger }.
\end{equation*}%
The corresponding matrix product wave functions for these excitations are
\begin{equation}
\left| \psi _{k}^{a}\right\rangle =\sum_{\{m_{i}\}}\mathrm{Tr}%
(B^{[m_{1}]}\cdots C_{a}^{[m_{k}]}\cdots B^{[m_{L}]})|m_{1}\cdots
m_{L}\rangle ,
\end{equation}%
where $C_{a}^{[m]}$ is defined by
\begin{eqnarray*}
&&\left(
\begin{array}{c}
C_{1}^{[m]}(p,q) \\
C_{0}^{[m]}(p,q) \\
C_{-1}^{[m]}(p,q)%
\end{array}%
\right) =\frac{1}{S}(-1)^{S-p}\sqrt{(S+m)!(S-m)!} \\
&&\text{ \ }\times \sqrt{\binom{S}{p-1}\binom{S}{q-1}}\left(
\begin{array}{c}
(S-q+1)\delta _{m,q-p+1} \\
(S-2q+2)\delta _{m,q-p} \\
(1-q)\delta _{m,q-p-1}%
\end{array}%
\right) .
\end{eqnarray*}

For one crackion in the string, it can be shown that the SOP in the
thermodynamic limit is given by
\begin{equation}
\lim_{\left| j-i\right| \rightarrow \infty }\langle \left( A_{i}^{z}\right)
^{\dagger }\prod_{k=i}^{j-1}e^{i\theta S_{k}^{z}}A_{j}^{z}\rangle _{\mathrm{%
cr}}=\mathcal{O}_{A}^{z}(\theta )\frac{\left\langle \lambda _{\max }^{%
\mathcal{O}}\right| G_{\mathcal{O}}^{a}\left| \lambda _{\max }^{\mathcal{O}%
}\right\rangle }{\left\langle \lambda _{\max }\right| G^{a}\left| \lambda
_{\max }\right\rangle },
\end{equation}%
where $\mathcal{O}_{A}^{z}(\theta )$ is the maximized string order parameter
without crackions. When replacing operator $\hat{P}$ by $\exp (i\theta
S^{z}) $ and $I$ in the following crackion transfer matrix:
\begin{equation}
G_{P}^{a}=\sum_{m,m^{\prime }}\left\langle m^{\prime }\right| \hat{P}\left|
m\right\rangle \left( \bar{C}_{a}^{[m^{\prime }]}\otimes C_{a}^{[m]}\right) ,
\end{equation}%
one obtains $G_{\mathcal{O}}^{a}$ and $G^{a}$, respectively. From these
definitions, one can show that the ratio between $\left\langle \lambda
_{\max }^{\mathcal{O}}\right| G_{\mathcal{O}}^{a}\left| \lambda _{\max }^{%
\mathcal{O}}\right\rangle $ and $\left\langle \lambda _{\max }\right|
G^{a}\left| \lambda _{\max }\right\rangle $ is equal to $\exp (ia\theta )$ ($%
a=\pm 1,0$) for the three kinds of crackions. If there are a few diluted
crackions between $i$ and $j$, the above result can be extended to
\begin{equation}
\lim_{\left| j-i\right| \rightarrow \infty }\langle \left( A_{i}^{z}\right)
^{\dagger }\prod_{k=i}^{j-1}e^{i\theta S_{k}^{z}}A_{j}^{z}\rangle _{\mathrm{%
\ cr}}=\mathcal{O}_{A}^{z}(\theta )\exp (i\theta \sum_{k=i}^{j}a_{k}),
\end{equation}%
where the phase factor $\exp (i\theta \sum_{k=i}^{j}a_{k})$ counts the total
number of crackions between $i$ and $j$.

For the $S=1$ VBS state, we have $\theta =\pi $ and the SOP alternates its
sign according to the parity of $\sum_{k=i}^{j}a_{k}$, which can be
interpreted as a $Z_{2}$ topological charge \cite{Suzuki-1995}. If we
consider the SOP in the $x$-direction, a crackion with $a=0$ also flips the
sign. In a\ general spin-$S$ case, the spin-twist angle of the maximized SOP
is given by $\theta =2\pi n/(S+1)$ ($n=1,\ldots ,S$). Our optimized SOP
suggests that the crackion carries a $Z_{S+1}$ topological charge, in
consistent with the hidden $Z_{S+1}\times Z_{S+1}$ symmetry argument.

In summary, we have shown that the hidden order in the VBS states can be
characterized by the generalized den Nijs-Rommelse-type SOPs. The
maximization of these SOPs automatically leads to spin-twist angles
corresponding to $Z_{S+1}$ rotations around $z$ or $x$ axes, suggesting the
existence of a hidden $Z_{S+1}\times Z_{S+1}$ symmetry. In the presence of
the crackion excitation, the maximized SOPs are shown to exhibit a $Z_{S+1}$
topological charge. Recently, it was shown that the den Nijs-Rommelse SOP is
an effective measure of the localizable entanglement in the $S=1$ VBS state %
\cite{Verstraete-Roncaglia}. We believe that our maximized SOPs provide a
natural extension of the den Nijs-Rommelse SOP and can be used to explore
multipartite entanglement properties of arbitrary VBS states.

We thank D. H. Lee and X. G. Wen for stimulating discussions. Support from
NSF-China and the National Program for Basic Research of MOST-China is
acknowledged.

\end{document}